\documentclass[preprint,showpacs,preprintnumbers,amsmath,amssymb]{revtex4}

\usepackage{graphicx}
\usepackage{dcolumn}
\usepackage{bm}

\begin{document}

\preprint{APS/123-QED}

\title{The Karlsruhe Astrophysical Database of Nucleosynthesis in Stars Project - \\
       Status and Prospects}

\author{I. Dillmann}
       \email[Corresponding author: ]{i.dillmann@gsi.de}
       \affiliation{$II.$ Physikalisches Institut, Justus-Liebig-Universit\"at Giessen, Germany and \\
       GSI Helmholtzzentrum f\"ur Schwerionenforschung Darmstadt, Germany}
       
       \author{T. Sz\"ucs}
       \affiliation{Institute for Nuclear Research (MTA Atomki), Debrecen, Hungary} 
       
       \author{R. Plag}
       \affiliation{Institut f\"ur Angewandte Physik, Goethe-Universit\"at Frankfurt, Germany and \\
      GSI Helmholtzzentrum f\"ur Schwerionenforschung Darmstadt, Germany}
       
       \author{Z. F\"ul\"op} 
       \affiliation{Institute for Nuclear Research (MTA Atomki), Debrecen, Hungary} 
       
       \author{F. K\"appeler} 
       \affiliation{Karlsruhe Institute of Technology (KIT), Institut f\"ur Kernphysik, Campus Nord, Karlsruhe, Germany} 
       
       \author{A. Mengoni} 
       \affiliation{Agenzia Nazionale per le Nuove Tecnologie, l'Energia e lo Sviluppo Economico Sostenibile (ENEA), Bologna, Italy} 
       
       \author{T. Rauscher} 
       \affiliation{Centre for Astrophysics Research, School of Physics, Astronomy and Mathematics, University of Hertfordshire, UK and \\Institute for Nuclear Research (MTA Atomki), Debrecen, Hungary and \\
       Department of Physics, University of Basel, Switzerland} 

\date{\today}

\begin{abstract}
The KADoNiS (Karlsruhe Astrophysical Database of Nucleosynthesis in Stars) project is an astrophysical online database for cross sections relevant for nucleosynthesis in the $s$ process and the $\gamma$ process. The $s$-process database (www.kadonis.org) was started in 2005 and is presently facing its 4th update (KADoNiS v1.0). The $\gamma$-process database (KADoNiS-p, www.kadonis.org/pprocess) was recently revised and re-launched in March 2013.

Both databases are compilations for experimental cross sections with relevance to heavy ion nucleosynthesis. For the $s$ process recommended Maxwellian averaged cross sections for $kT$= 5-100~keV are given for more than 360 isotopes between $^{1}$H and $^{210}$Bi. For the $\gamma$-process database all available experimental data from $(p,\gamma), (p,n), (p,\alpha), (\alpha,\gamma), (\alpha,n)$, and $(\alpha,p)$ reactions between $^{70}$Ge and $^{209}$Bi in or close to the respective Gamow window were collected and can be compared to theoretical predictions. The aim of both databases is a quick and user-friendly access to the available data in the astrophysically relevant energy regions. 
\end{abstract}

\maketitle

\section{\label{intro}Introduction}
The nuclei heavier than iron are produced in stars by the three processes called $s$-, $r$-, and $p$-process \cite{BBFH}. The first two involve neutron capture reactions and $\beta$-decays. The main difference is the neutron density and therefore the time scale of the neutron capture. For the $s$-process the capture time scale is much longer (typically one capture per year) than the $\beta$-decay half-lives, therefore the reaction flow runs along the valley of stability up to $^{209}$Bi. In the $r$-process the neutron density is so high ($\gg$10$^{20}$ cm$^{-3}$) that subsequent neutron captures drive the material far away from the line of stability close to the neutron drip-line. When the neutron density drops and the temperature decreases (``freeze out"), the material decays back to stability via long $\beta$-decay chains. 
       
For about 30 proton-rich isotopes between $^{74}$Se and $^{196}$Hg other production mechanisms involving high temperatures and charged particles are required. These so-called ``$p$-nuclei" are responsible for only less than 1\% of the total solar abundance in mass range above iron, and are mainly produced in a mechanism called ``$\gamma$ process".

\subsection{Nuclear Data Needs for the $S$ Process}
       
       The nuclear data needs for all heavy ion production processes involve mainly cross sections and half-lives of the participating nuclei. In the stellar $s$-process environment temperatures of up to several hundred million K are attained. This requires the knowledge of cross sections in the energy range of a few tens of eV up to a few hundred keV for the important reactions, which are mainly $(n,\gamma)$ but also some $(n,p)$ and $(n,\alpha)$ reactions. However, the involved nuclei can also exist in excited states, whereas the measurements in the laboratory include only nuclei in the ground-state (with exception of the quasi-stable $^{180}$Ta$^m$), and thus an additional correction is needed.

       Also the stellar decay rates may be different from terrestrial ones, due to thermal excitation of the nuclei and, in the case of electron capture, due to ionization. Special cases with respect to $s$-process nucleosynthesis are $^{163}$Dy and long-lived radioactive isotopes like e.g. $^{79}$Se. $^{163}$Dy is stable under terrestrial conditions, but can undergo bound-state $\beta$-decay when it is fully ionized ($^{163}$Dy$^{66+}$ has a half-life of 47~d \cite{Jung92}). Among the nuclei involved, the long-lived isotopes like $^{63}$Ni and $^{79}$Se assume key positions, because their stellar $\beta$$^-$-decay rate becomes comparable to the neutron capture rate ($\lambda_\beta$$\approx$$\lambda_n$). The resulting competition leads to branchings in the $s$-process nucleosynthesis path. These branching-point isotopes can be used for an in-situ diagnosis of the temperature and neutron density. 
       
       It was noticed that the effective branching ratios for $^{79}$Se and $^{85}$Kr are almost equal although their terrestrial half-lives differ by more than four orders of magnitude. The solution are $\beta$-decays from thermally populated excited states. For $^{79}$Se the $\beta$-decay rate is almost constant for $T$$<$10$^8$K, but drops quickly to the order of a few years when low-lying first excited states become thermally populated \cite{KlK88}. Since the temperature dependence of the half-life is well known from calculations \cite{KlK88,TaY83}, the branching at $^{79}$Se can be interpreted as an $s$-process thermometer.
       
       Another important ingredient for a reliable $s$-process simulation are accurate cross sections. In the weak $s$-process ($A$=60-90) uncertainties of single cross sections can produce a long-range propagation effect which affects the whole reaction flow up to the next shell closure at $N$=50. The reason is that no local equilibrium exists in the weak component (the product abundance times cross section for neighboring isotopes in the reaction path is no longer approximately the same). This was first demonstrated for $^{62}$Ni(n,$\gamma$) \cite{HN05} where until five years ago experimental values at stellar temperatures ($kT$=30 keV) ranged between 12 and 37 mb. The higher cross section lead to a 45\% higher yield of all isotopes between $A$=65 and 90 in stellar models compared to the previously recommended value of 12~mb. A careful analysis including new measurements have led to a new recommended value of 22.3$\pm$1.6~mb at $kT$=30~keV (see KADoNiS v0.3, www.kadonis.org), which is also in good agreement with the most recent n\_TOF result \cite{CL13a}. Also for the neighboring radioactive branching isotope $^{63}$Ni first experimental results are now published \cite{CL13a,CL13b} which lead to a factor of 2 higher cross section at $kT$= 30~keV than previously recommended.
       
       Other important nuclei in the $s$ process, where accurate cross sections are indispensable, are isotopes at the neutron shell closures and the so-called "$s$-only isotopes". The latter are of pure $s$-process origin and thus represent the most direct probes of the $s$-process abundance distribution. The desired uncertainty at stellar temperatures is $\pm$1\% which is up to now only achieved for some isotopes between $A$=110-176, the average uncertainty at $kT$=30~keV is $\pm$3.8\%.
       
       The isotopes at the shell closures are directly responsible for the peaks in the solar $s$-abundance distribution because their small capture cross sections are hindering the reaction flux towards higher masses. The average uncertainty at $kT$=30~keV for all 19 stable isotopes at $N$=28, 50, 82, and 126 is presently $\pm$8.3\%.
       
       Another important group of nuclei are the so-called ``neutron poisons". These can be all isotopes with large abundances and/or neutron capture cross sections which reduce the available neutrons per seed nuclei. In massive stars the reaction $^{14}$N$(n,p)$$^{14}$C has the largest influence on the neutron economy. Other prominent cases are $^{12}$C and $^{16}$O due to their large abundances. In the case of $^{12}$C the neutrons are recovered via $^{12}$C$(n,\gamma)$$^{13}$C$(\alpha,n)$$^{16}$O. However, the large abundance of $^{16}$O makes it again to a neutron poison ("self-poisoning"). A similar behavior can be seen also with the second neutron production reaction, $^{22}$Ne$(\alpha,n)$$^{25}$Mg. Since $^{25}$Mg is produced in large amounts, part of the newly created neutrons are immediately consumed by the $^{25}$Mg$(n,\gamma)$ reaction. All three capture reactions exhibit up to now rather large uncertainties in the order of 10\%.
       
       Last but not least the neutron source reactions itself, $^{13}$C$(\alpha,n)$$^{16}$O and $^{22}$Ne$(\alpha,n)$$^{25}$Mg, need improvements. Despite all efforts in the last years these reactions are still rather uncertain in the relevant energy regions. For a detailed discussion, see \cite{WKL12}.
       
       \subsection{Nuclear Data Needs for the $\gamma$ Process}
       For the $\gamma$ process at temperatures of $T$= 2$-$3~GK photon-induced reactions on pre-existing seed nuclei dominate. Detailed reaction network calculations involve thousands of nuclei and ten thousands of reactions, mostly on unstable nuclei. Unfortunately, these calculations still fail to reproduce the observed solar abundance for all $p$-nuclei within a factor of 3 \cite{AG03,RDD13}, especially for the most abundant nuclei $^{92,94}$Mo around $N$=50.
       
       The nuclear physics input for these calculations comes mainly from Hauser-Feshbach model calculations. To test these predictions systematic measurements in the relevant mass and energy range are needed. Experimentally it is easier to investigate the inverse radiative capture reactions and to calculate the photo-induced reaction rate by applying the principle of detailed balance \cite{TR09,TR11}. Additionally, the experimental ground-state rate for capture provides a larger contribution to the stellar rate than for photodisintegration (with a few exceptions, such as $^{148}$Gd($\gamma,\alpha$)$^{144}$Sm).
       
       The improvement of the predictive power for reaction cross sections is crucial for further progress in $p$-process models, either by directly replacing theoretical predictions by experimental data or by testing and improving the reliability of statistical models, if the relevant energy range, the so-called Gamow window, is not accessible by experiments. 
       
       Until recently no dedicated database for the $p$ process existed, where $(p,\gamma), (p,n), (p,\alpha), (\alpha,\gamma), (\alpha,n)$, and $(\alpha,p)$ reactions for isotopes between $^{70}$Ge and $^{209}$Bi within the respective Gamow window were collected. This energy range is typically between 1-6~MeV for proton-induced and 3.5-13.5~MeV for $\alpha$-induced reactions.

       \section{The KADoNiS project}
       \subsection{History of Stellar Neutron Capture Compilations}
       The pioneering compilation for stellar neutron capture cross sections was published in 1971 by Allen and co-workers \cite{AGM71}. This paper reviewed the role of neutron capture reactions in the nucleosynthesis of heavy elements and presented also a table of recommended (experimental or semi-empirical) Maxwellian averaged cross sections (MACS) at $kT$= 30~keV (MACS30) for nuclei between carbon and plutonium.
       
       The idea of an experimental and theoretical stellar neutron cross section database was picked up by Bao and K\"appeler \cite{bao87} in 1987 for $s$-process studies. This compilation included cross sections for $(n,\gamma)$ reactions between $^{12}$C and $^{209}$Bi, some $(n,p)$ and $(n,\alpha)$ reactions for isotopes between $^{33}$S and $^{59}$Ni, and also (n,$\gamma$) and (n,f) reactions for long-lived actinides. A follow-up compilation was published in 1992 \cite{BVW92}, and in 2000 the compilation \cite{bao00} was extended to Big Bang nucleosynthesis. It now includes a collection of recommended MACS30 values for isotopes between $^{1}$H and $^{209}$Bi, and semi-empirical recommended values for nuclides without experimental cross section information. These estimated values are normalized cross sections derived with the Hauser-Feshbach code NON-SMOKER \cite{NS}, which account for known systematic deficiencies in the nuclear physics input. Additionally, the database provided stellar enhancement factors and energy-dependent MACS for thermal energies between $kT$= 5 keV and 100 keV.
       
       The first version of the "Karlsruhe Astrophysical Database of Nucleosynthesis in Stars" (KADoNiS) was released in 2005 \cite{kadonis}. This online database was an updated sequel to the previous compilations. KADoNiS was updated three times until now, the fourth update is in progress. The current version v0.3 \cite{k03} contains recommended MACS values at $kT$=30~keV for 359~isotopes, out of which 80 are semi-empirical estimates based on theoretical data and experimental values of nearby isotopes. The average uncertainties for the (experimental) MACS at $kT$= 30~keV are $\pm$6.5\%. Experimental information exists for almost all stable isotopes with exception of $^{17}$O, $^{36,38}$Ar, $^{40}$K, $^{50}$V, $^{70}$Zn, $^{72,73}$Ge, $^{77,82}$Se, $^{98,99}$Ru, $^{131}$Xe, $^{138}$La, $^{158}$Dy, and $^{195}$Pt.
       
       The $s$-process database of KADoNiS provides not only recommended values but lists also the available experimental data as well as additional information from some evaluated databases, namely ENDF/B, JENDL, and JEFF. The latter data can also be retrieved from www.nndc.bnl.gov/astro/ \cite{PM12}

       \subsection{KADoNiS v1.0}
       In the present update, the previously missing datasets for $^{6}$Li, $^{10,11}$B, and $^{17}$O were included, increasing the number of isotopes to 363. From the above mentioned 16 stable isotopes without experimental information in the stellar energy range, new measurements on $^{50}$V, $^{70}$Zn, $^{72,73}$Ge, $^{77,82}$Se, and $^{195}$Pt were included.
       
       A further extension to more neutron-rich isotopes (2-3 mass units away from stability) is under discussion, but only semi-empirical estimates will be available for these cases. The values from most recent evaluated libraries ENDF/B-VII.1 \cite{endfb71} and JENDL-4.0 \cite{jendl40} were included, and were used in many cases instead of older time-of-flight measurements. 
       
       As mentioned before, only reactions on nuclei in their ground states (g.s.) are measured in laboratory experiments. In the stellar environment, however, also reactions on thermally excited states of the nuclei contribute to the reaction rate. In the past, experimental g.s. MACS and g.s. rates have been converted to stellar ones using a stellar
       enhancement factor (SEF), given by the theoretically predicted ratio between stellar rate and g.s. rate. This approach has been found to be inadequate \cite{RMD11} and will be replaced by the procedure given in \cite{Rau12}, which also allows to properly define uncertainties in the stellar rates from a combination of theoretical and experimental uncertainties. 
       
       
       Once the update with new experimental data is complete, semi-empirical values for those isotopes without experimental information will be recalculated as in \cite{bao00}. The new data will then be included in state-of-the-art $s$-process models to investigate the influence of changes on the calculated $s$-abundance distribution.
       
       \begin{figure*}[!htb]
       \includegraphics[width=0.85\textwidth]{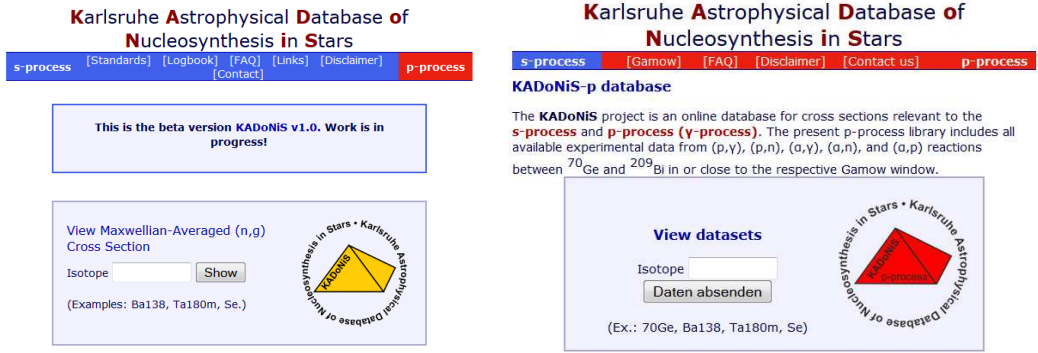}
       \caption{Web appearance of the $s$-process database (left) and the $\gamma$-process database (right) of KADoNiS at www.kadonis.org.}
       \label{fig1}
       \end{figure*}
       
       \subsection{KADoNiS-p: The astrophysical $P$-process Database}
       The aim of the $p$-process database is the collection of all available experimental cross sections in or close to the respective Gamow window and the comparison with various theoretical models. In 2006 a first draft for such a database within the KADoNiS project was launched, but it lasted until 2010 before this part was improved in collaboration with ATOMKI Debrecen, Hungary \cite{TS10,TS12}.
       
       The new $p$-process compilation started essentially from the EXFOR database \cite{exfor}, which has the advantage that it contains (almost) all available experimental cross sections and is regularly updated. EXFOR was carefully scanned for experimental data from $(p,\gamma), (p,n), (p,\alpha), (\alpha,\gamma), (\alpha,n)$, and $(\alpha,p)$ reactions for targets heavier than $^{70}$Ge and energies down to 1.5 times the upper end of the respective Gamow window for $T$= 3~GK. This upper cut-off energy was arbitrarily chosen because most available data were measured above the astrophysically relevant Gamow windows.
       
       In the present database 309 datasets for 181 different charged-particle reactions between $^{70}$Ge and $^{209}$Bi are available. The revised version of KADoNiS-p is available online at www.kadonis.org/pprocess (Fig.~\ref{fig1}) since March 2013. For a more detailed description, please refer to Ref.~\cite{TS13}.
       
       \section{ CONCLUSIONS}

       The KADoNiS project (www.kadonis.org) is providing astrophysical cross sections for $s$- and $p$-process nucleosynthesis. The main aim is to maintain user-friendly and easily accessible compilations of the available (experimental) data in the relevant energy ranges. Comparisons to some theoretical predictions are available, as well as to data from evaluated libraries. 
       
       
       This work is supported by the Helmholtz association via the Young Investigators projects VH-NG-627 and VH-NG-327, the Hungarian Scientific Research Funds OTKA NN83261 (EuroGENESIS) and OTKA K101328, the Hungarian Academy of Sciences, the MASCHE collaboration in the ESF EUROCORES program EuroGENESIS, the THEXO collaboration within the
       7th Framework Programme of the EU, and the Swiss National Science Foundation.
       

\end{document}